\documentclass[prb,showpacs,twocolumn]{revtex4}
\usepackage{graphicx}% Include figure files
\usepackage{dcolumn}% Align table columns on decimal point
\usepackage{bm}% bold mat
\usepackage{amsbsy}
\begin{document}
\title{Intrinsic spin-orbit interactions in flat  and curved graphene nanoribbons}
%\title{Symmetry-dependent spin-orbit splitting in carbon nanotubes}
%\title{Spin polarization in carbon nanotubes induced by spin-orbit interaction}
%
\author{
M. P. L\'opez-Sancho, and M.C. Mu\~noz}
\affiliation{ Instituto de Ciencia de Materiales de Madrid, Consejo Superior de
Investigaciones Cient{\'{\i}}ficas, Cantoblanco, 28049 Madrid, Spain}

\date{\today}
\pacs{73.20.-r, 73.22.-f, 73.23.-b}

\begin{abstract}
Recent theoretical and experimental works on carbon nanotubes and
graphene samples have revealed that spin-orbit interactions, though
customarily ignored in carbon-based materials, are more important
and complex than it was thought. We study the intrinsic spin-orbit coupling
effects on graphene nanoribbons, both flat and bent. Calculations
are performed within the tight-binding model with the inclusion of a
four-orbital basis set. Thereby the full symmetry of the honeycomb
lattice and the hybridization of $\sigma$ and $\pi$
bands are considered. In addition to the zero-energy $\pi$-edge states,
$\sigma$-derived edge states are found for  the three investigated 
ribbon geometries. The $\sigma$ states are also spin-filtered and
localized at the boundaries of the ribbons. The calculated spin-orbit splittings
are larger  for the $\sigma$- than for the $\pi$-derived edge states.
Due to this enhancement, spin-orbit splittings of the  $\sigma$-states
reach values of the order of a few Kelvin.
These spin-filtered edge states are robust under $\sigma-\pi$
hybridization and curvature effects. 

\end{abstract}

\maketitle

%%%%%%%%%%%%%%
\section{Introduction}

The influence that geometry and size have on the electronic
properties of carbon based materials has been already proved. Two
dimensional graphene sheets, graphene  stripes known as graphene
nanoribbons (GNRs), or carbon nanotubes (CNTs) formed by rolling a
graphene sheet unto a cylinder,  
due its unique electronic properties mainly to their geometries. 
Chirality and diameter determine the properties of CNTs, while the
GNRs physics depends on its width and the shape of
its edges \cite{NGPNG09}. 
This strong dependence of the electronic properties on
the geometry arises from the peculiar structure of graphene with
carbon atoms ordered in a honeycomb lattice, divided into two nonequivalent 
triangular sublattices, and confers these materials promising potential applications in
electronics and spintronics.
The low energy physics of carbon-based materials is governed by the states 
close to the Fermi energy in graphene.
They correspond to the $\pi$-states
near the nonequivalent $K$ and $K'$ points at opposite corners of the hexagonal
Brillouin zone (BZ),  the Dirac points, at which the valence and 
conduction bands touch and present a conical energy spectrum. 

Due to the low atomic number of carbon, spin-orbit interaction (SOI)
was expected to be small. Nevertheless, several 
theoretical studies were devoted to investigate SOI effects in graphite \cite{DD65}
and more recently in CNTs  and graphene \cite{A00,EM01,MEHB02,CLM04,HHGB06,MHSKMD06}. 
In a $\bf k.p$ scheme and using a perturbation approach, it was shown 
that the SOI opens a gap in the energy bands crossing at the Fermi level
and lifts the spin degeneracy for higher energy subbands \cite{A00}. In a previous
work \cite{CLM04}, within a four-orbital tight-binding (TB) framework, we 
have demonstrated the 
chiral dependence of the intrinsic SOI effects, which can not  be described 
by a continuum model. Spin-splitting is only possible for chiral CNTs 
due to the lack of inversion symmetry, while the spin-degeneracy remains 
for achiral CNTs. Furthermore, the recently measured 
asymmetric splittings of valence and conduction bands in a CNT quantum dot \cite{KIRM08} 
are  also explained within the same model \cite{CLM09,JL09,ISS09} pointing
out that curvature  enhances the SOI strength. In fact, 
recent experimental  measurements of  the energy shifts caused by the SOI,
have proved that the SOI effects in CNT
quantum dots and graphene samples are not as small as  predicted \cite{KIRM08,GDM10}. 
Besides, since SOI is the key ingredient for the Quantum Spin
Hall state, the new phase of matter proposed by Kane and Mele \cite{KM05} in graphene,
the previously considered almost negligible SOI interaction in graphene-based 
nanostructures has risen to be
a main topic in the condensed matter field \cite{ZS07,ZBS08,ZS09,WHZ09,GM10,SL10}. 

The electronic properties of GNRs have been intensively investigated due to
the remarkable behavior of the zero-modes edge 
states \cite{FWNK96,RH02,PCNG06,BF06,RK08,WTYS09,WOTFN10}. Most of the works
have been based either on the $\bf k.p$ or the effective $\pi$-orbital TB approaches, 
the simplest models that capture the physics of graphene, although  
some first-principles and LCAO \cite{SCL06,BT07,ZY09} calculations have also been reported.
Furthermore, the effect of the SOI in graphene sheets and GNRs with 
different boundaries is currently attracting
great interest and, in spite of the large amount of published works, many
aspects are still hidden. Some numerical calculations based
on multiorbital models \cite{OIKI08}, and first principles \cite{PCMH07,GKEAF09} can be
found in the literature but, to our knowledge,
no investigation of the SOI on flat and bent GNRs based on a $sp^3$ tight-binding  model 
has so far been performed.

We address here the
study of the intrinsic spin-orbit coupling effects on the electronic
properties of GNRs with edges of
different shapes: zigzag, Klein's bearded and armchair edges. We consider
flat and bended ribbons to investigate the interplay between the SOI
and the curved geometry and focus on the topological edge states,
not only the well known $\pi$-derived states, but also those arising from 
$\sigma$-orbitals. We investigate how curvature influences the interplay of confinement and 
SOI in determining the topological character of the edge states.

An explicit description of edge states requires a model that gives
the energy bands through the entire BZ. Therefore, calculations are performed 
by exactly solving an
empirical tight-binding Hamiltonian with a $sp^3$ basis set. 
The TB model handles equally all points in the BZ, which is not possible with
the continuum model and is especially appropriated with regard to the treatment
of the SOI. Besides,  it takes into account the full
symmetry of the honeycomb lattice, hence the threefold rotational symmetry and
the trigonal warping are included. Nor external potentials or doping are 
taken into account, although the effect of the $\sigma$-derived
edge states on the behavior of the spin Hall conductivity with the chemical 
potential would be discussed.

In addition to the $\pi$-edge states we found that
$\sigma$-derived edge states are also present for all the boundary geometries of 
the GNRs studied. Further, as expected its energy is independent of GNR width and 
for a large interval of {\bf k} values  of the one dimensional BZ the $\sigma$ edge sates are 
located in a gap. Hence, they are likewise perfectly localized
along the edges of the ribbon.
Our main finding is that, when the SOI term is included in
the Hamiltonian, independent of the GNR termination, the fourfold degeneracy of edge states, 
both of $\pi$  and $\sigma$ character, is lifted
except at the time reversal invariant symmetry  points $\Gamma$ and $K$ of the one dimensional BZ.
They split into the spin filtered edge-states, which 
are robust upon $\sigma-\pi$ bands
hybridization and, as expected for a topological state, are unaffected by curvature.

The paper is organized as follows: in Section II the details of the
model are given. In Section III the results obtained for the different type of GNRs are 
described, and curvature effects are analysed by comparison of
flat and curved geometries. Finally, we conclude in Section IV summarizing
our results and drawing some conclusions.

\section{Model and Method}

\begin{figure}
\leavevmode
\includegraphics[width=\columnwidth]{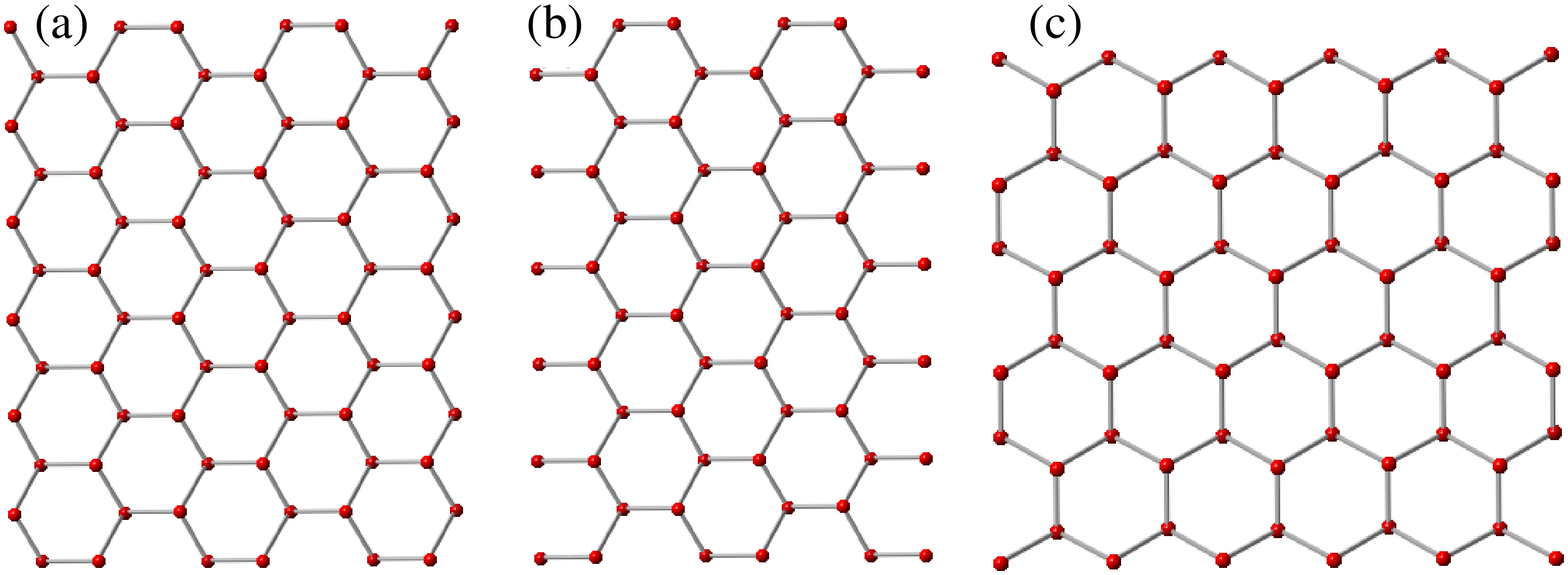}
\caption{ 
Schematic lattice structure of graphene nanoribbons with: a)zigzag, b) bearded, and
c) armchair edges.}
\label{struc1}
\end{figure}

A ribbon consists of a stripe of graphene of infinite length 
and finite width. There are two prototypical shapes of the GNR edges, zigzag
and armchair, they have a $\theta= 30º$ angle of difference in the cutting 
direction and the termination of a generic ribbon has a combination of the two types.
We consider ribbons of three different kinds of terminations: zigzag, bearded and armchair,
and define the width of the GNRs as {\it n}, where {\it n} stands for the number of 
zigzag lines for the zigzag and bearded ribbons and  for the number of dimer 
lines for the armchair ribbon \cite{FWNK96}. The corresponding 
unit cell contains ${\it N}$ = 2${\it n}$ carbon atoms. 
Examples of the three types of  ribbons are displayed in Figure~\ref{struc1}.
We do not consider here any reconstruction
or relaxation of the edges\cite{JHT09,GMCL09}. 
Nevertheless, chemical and structural modifications 
of graphene edges are promising routes to design GNR based devices\cite {RK08,WOTFN10,SL10}.

Curvature effects
are introduced by bending the ribbon along its width. The bending
is realized without stretching, varying the atom coordinates
to form an open cylinder. Different curvatures are obtained
by changing the angles and diameters of the cylindrical
configurations. 

The electronic properties of the graphene nanoribbons are calculated from the  
one-electron Hamiltonian given by,

$$H = H_{0} + H_{SO},$$

\noindent
$H_{0}$ is the spin-independent noninteracting Hamiltonian,

$$H_{0} = \frac{\bf p^2}{2m}  + V({\bf r}),$$

\noindent
and and $H_{SO}$ is the microscopic spin-orbit interaction term

$$H_{SO} = \frac{\hbar}{4m^2c^2} {\bf S}\cdot{({\bf \nabla} V \times {\bf p})},$$

%$H_{SO} = \frac{\hbar}{4(mc^2} {\bf \sigma} (\nabla V \times {\bf p})$,

\noindent
where V is the full crystal potential, ${\bf p}$ the
electron momentum and ${\bf S}$ represents the spin operator.

We calculate the band structure of the graphene ribbons by the Slater-Koster \cite{SK} empirical 
tight-binding (ETB) Hamiltonian.
A four-orbital, $2s$, $2p_x$, $2p_y$ and $2p_z$, basis set is considered in order 
to include the conventional on-site approach for the intrinsic SO interaction.
Within the TB approximation the
$H_{0}$ term is written as,

\begin{equation}
H= \sum_{i,\alpha,s} \epsilon_{\alpha}+\sum_{<ij>,\beta,s} t_{ij}^{\alpha,\beta} c_ {i, s}^{\alpha+}c_{j,s}^{\beta}+ H.c. ,
\end{equation}

\noindent
where $\epsilon_\alpha$ represents  the atomic energy of the
orbital $\alpha$, $<ij>$ stands for atomic sites of the honeycomb lattice
and $c_{i, s}^{\alpha+}$ and $c_{i, s}^{\alpha}$ are the 
creation and annihilation operators of one electron at site $i$, 
orbital $\alpha$ and spin $s$, respectively.

\noindent

Since the major contribution of the crystal potential, $\nabla V$, is near the
atomic nuclei, the intrinsic SOI can be accurately approximated by a local atomic 
term of the form: 

$${ H_{SO}} = \frac{\hbar}{4m^2c^2} \frac{1}{r_i} \frac{d V_i}{d r_i} {\bf L}\cdot{\bf  S} = \lambda {\bf L}\cdot{\bf S}, $$

\noindent
where spherical symmetry of the atomic potential is assumed. $r_i$ is the radial 
coordinate with origin at the atom, $V_i$ the spherical symmetric potential
about the same atom and
${\bf L}$ the orbital angular momentum. The atomic 
SO coupling constant $\lambda$ depends on the orbital angular momentum ${\bf L}$ and 
thus on the atomic orbital. 

For the $sp^3$ model Hamiltonian, SO coupling 
occurs only among {\it p} orbitals and neglecting
nearest neighbor SO terms, the
intrinsic SOI is described as an on-site interaction among the {\it p-}orbitals. 
%$H_{SO}$ term couples $p$ orbitals on the same atom. 
%This approach is reasonable, due to the small atomic number of carbon atoms. 
In contrast,
in the {\it \bf k.p} and effective $\pi$-band models, the atomic SOI is treated
by perturbation theory as a second-neighbor spin dependent hopping term\cite{A00,HHGB06}

${ H_{SO}}$ adds
diagonal
and off-diagonal spin-dependent matrix elements to the $8N \times 8N$
matrix Hamiltonian, where N is the number of carbon atoms of the
unit cell and 8 comes from the four-orbital per spin TB basis set. 
Using the raising and lowering angular momentum operators,
 $L_+=L_x+iL_y$ and $L_- =L_x-iL_y$, respectively and the Pauli spin matrices,
%$\sigma$.  
the complete Hamiltonian, H,
in the $2 \times 2$ block spinor structure is given by,

\begin{equation}
H = \left(
\begin{array}{cc}
H_0 + \lambda L_z  &   \lambda L_-\\
\lambda L_+  &    H_0 - \lambda L_z   \\
\end{array}
\right).
\end{equation}

\noindent
where the atomic-like spin-orbit term  ${ H_{SO}}$ has been added to 
SK-ETB spin-independent $H_0$ Hamiltonian. The diagonal terms act as
an effective Zeeman field producing gaps of opposite signs at the 
$K$ and $K'$ points of the BZ.

In the $sp^3$ TB model the non-vanishing matrix elements of ${ H_{SO}}$
couple $2p_x$, $2p_y$ and $2p_z$ orbitals. These terms have been widely discussed
previously, see for example Ref.\cite{CH77,GM99}. 
${ H_{SO}}$ induces $\sigma-\pi$ hybridization and  does not break 
the time reversal symmetry, therefore  spin degeneracy can not be removed
on systems with inversion symmetry.

\begin{figure}

\includegraphics[width=7.0cm,clip]{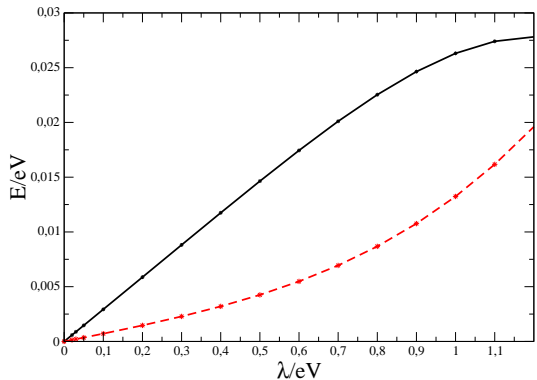}
\caption{ (Color on line) Energy splitting for the conduction (circles)
and valence (diamonds) bands of the (61,0) zigzag CNT as a function of
the value of the spin-orbit coupling constant $\lambda$. The
solid(black), dashed(red) lines are a guide to the eyes.}
\label{landa}

\end{figure}

The electronic properties of the GNRs are obtained
by the exact diagonalization of the total Hamiltonian $H$.
The To\-m\'a\-nek-Louie
parametrization for graphite \cite{TL}, previously used to calculate SOI
effects on CNTs, has been also employed in the present calculations.
The TB parameters, taken from Ref\cite{TL} are :  
 $\epsilon_s$= -7.3$eV$, $\epsilon_p$= 0.0$eV$,  $ss\sigma$= -4.30$eV$,
 $sp\sigma$ = 4.98$eV$,  $pp\sigma$= 6.38$eV$,  $pp\pi$=-2.66$eV$,        
and  specify the  energy scale of the model.

The spin-orbit coupling strength  of graphene is unknown and its
exact value is under debate.
Due to the
small atomic number of carbon, it has been assumed to be very small and  different values
in the order of $\approx meV$ have been given \cite{KM05,HHGB06,MHSKMD06}.
In order to have a quantitative estimate of the SOI induced splittings,
we have evaluated $\lambda$ from the  
SO induced gaps experimentally measured in a CNT quantum dot \cite{KIRM08}.
In the experiment a zigzag small gap CNT of about $5$ $nm$ of diameter is probed
using tunnelling spectroscopy. The anisotropic
 gaps  measured at zero magnetic field
are of $\approx 0.37$ $meV$ for electrons and of $\approx 0.22$ $meV$ for holes.
The (61,0) CNT with a diameter of about $4.8$ $nm$ is a zigzag semiconducting tube of
the $n=3m+1$ family, so the SOI induced splitting
is greater for the conduction than for the valence band \cite{CLM09} as occurs
in the experiment. Calculated SOI splittings for the conduction and
valence bands are represented in
Fig.\ref {landa} 
for the (61,0) nanotube as a function of $\lambda$. 
From the figure it is inferred that 
 $\lambda \approx 3$ $meV$ is needed in  order to
to obtain SOI splittings  similar to those experimentally reported. This 
value is of the same order of magnitude of previous estimationsi \cite{MHSKMD06,HHGB06}. 
In the figures we present results obtained
with values of $\lambda$  in the range of $0.2-0.4 eV$,  for
illustrative purposes. This range  corresponds to $0.03-0.06 $ in 
units of the $pp\sigma$=6.38 $eV$  parameter.
% will be used in the figures for the sake of clarity\cite{OIKI08}. 
%The curved geometry of the tube enhances the
%spin orbit coupling, which is inversely proportional to the radius of the tubes
%\cite{CLM09}, but the diameter of the  (61,0) nanotube is is big enough and
%$the diameter is much greater than the interatomic distancea/D \approx 2.46/47.76= 0.0515$*****
% of the system therefore, 
%$E_{{\bf k}, \sigma} =E_{{-\bf k}, -\sigma} $
%where $E_{{\bf k}, \sigma}$ is the energy of the eigenstate with
%wavevector ${\bf k}$ and spin $\sigma$.  If the crystal also has
%inversion symmetry, that is,
%$E_{{\bf k}, \sigma} =E_{{-\bf k}, \sigma}$, then
%spin degeneracy cannot be removed,
%$E_{{\bf k}, \sigma} =E_{{\bf k}, -\sigma}$.
%Edge states appear when truncation of inter-atomic bonds
%occurs due to the presence of boundaries in the system.
%The inclusion of $\sigma$ orbitals  adds new bands to the structure
%and, therefore, when matrix elements of the Hamiltonian are set at zero
%to create the boundaries, edge states will originate in the
%$\sigma$ related dangling bonds. 
%The SOI effects are investigated in the zero-energy edge states 
%derived from the $\pi$ orbitals and in the edge states due to the
%presence of $\sigma$ orbitals.   

\section{Results}

The electronic properties of GNRs are derived from the band structure
of graphene subject to a
stripe geometry. The combination of the confinement due to the finite-size
and the presence of boundaries yields the peculiar band structure of
GNRs. The truncation
of inter-atomic bonds caused by the borders gives rise to the appearance of
edge states, which are
strongly dependent on the atomic termination of the GNR. Moreover,
the energy subbands associated to the intrinsic band structure of the
graphene sheet are also dependent on the boundary conditions of the GNR.
Calculations have been carried out in GNRs of different geometries
and widths. In order to avoid the coupling between edge states
localized at each boundary, ribbons of more than 50 chains
are considered. 

\subsection{Zigzag graphene nanoribbons}

In zigzag graphene nanoribbons (ZGNRs) the atoms of each edge belong
to the same sublattice and opposite edge atoms are of
different sublattices. In the absence of SOI, the band structure of ZGNRs 
presents together with the subbands originating from the two-dimensional 
graphene electronic 
structure, zero-energy $\pi$-orbital derived states at the 
$\frac{ 2}{3}\pi <  k < \pi$ interval. These states, with wave functions 
localized at the edges, form flat bands and give rise to 
a peak in the density of states
at the Fermi energy being crucial for magnetic instabilities \cite{FWNK96,VLSG05}.
Besides, between $1$ and $3 eV$ below the Fermi energy, $\sigma$-orbital derived edge band 
appears in the $sp^3$ TB calculation. These states, which are missed in 
one-$\pi$ band models, disperse along the BZ and their
wave function amplitude is also fully localized at the edge atoms around $\bf k=0$ and 
$\bf k=\pm \pi$, where the ZGNRs have energy gaps. 
It is worth mentioning that ribbons of different
widths have been studied, from $n=10$, to $n=200$ and 
edge states, both $\pi$ and $\sigma$, appear at similar  energies   
independent of the ribbon width. 

\begin{figure}

\includegraphics[width=7.0cm,clip]{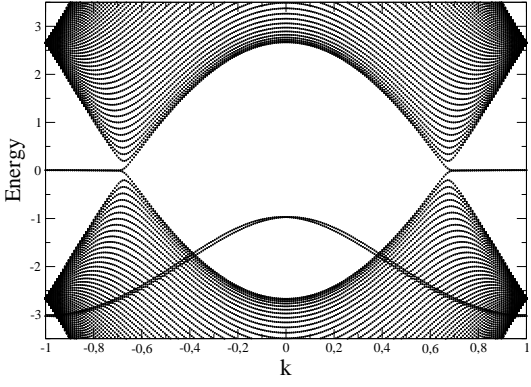}
\caption{Band structure of the zigzag ribbon n=60 obtained with SOI $\lambda=0.4$.
$k$ are in units of $\pi$}
\label{zzr30}

\end{figure}

%$-\frac{ 2\pi}{3}\leq {\abs{\bf k}}\leq  \frac{ 2\pi}{3}$. 
%The main features
%are analogous to those of the {\it n} = 10 ZGNR, although due to the larger 
%width there are in fact more subbands derived from the intrinsic band structure of
%graphene. 
%They
%are located at zero energy, as has been repeatedly predicted for the gapless
%$\pi$ edge states
%in previous calculations \cite{FWNK96,RH02,E06,BF06,PCNG06} and around 1 eV at 
%${\abs{\bf k}}$ = 0 for the $\sigma$ band.

\begin{figure}
\leavevmode
\includegraphics[width=7.0cm,clip]{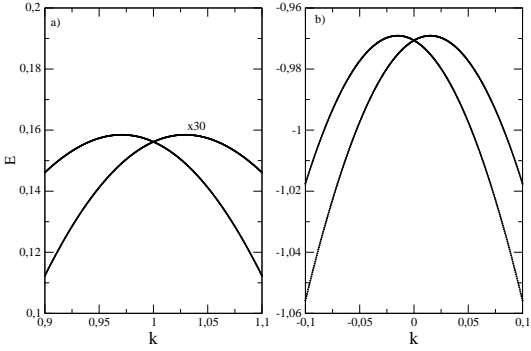}
\caption{Zoom of the band structure of the ribbon N=60 with $\lambda=0.4$
for the two edge states, left  around $k=1$ for the zero edge states
(energies are multiplied by 30)  and right in the region of $k=0$ for
the $\sigma$ derived band. $k$ are in units of $\pi$}
\label{zedst}
\end{figure}

To analyze the result of the SOI in ZGNRs,
the band structure of the {\it n} = 60 flat ZGNR with $\lambda$ = $0.4eV$ is
shown in Figure \ref{zzr30} in the range of $-\pi \leq {\bf k} \leq \pi$. The spectrum remains
gapless and the degeneracy of edge sates, originated from both $\pi$- 
and $\sigma$- orbitals, is partially lifted. 
A zoom of the dispersion relation of the $\pi$  and $\sigma$ 
edge states around the time-reversal invariant 
$\bf k=\pm \pi$  and $\Gamma$ points, respectively, are presented in Figure \ref{zedst}.
The SOI shifts down in energy the edge states giving
a small dispersion to the $\pi$ flat bands. It also lifts partially the fourfold 
degeneracy of both $\pi$ and $\sigma$ edge states, which become twofold 
degenerate, except at the ${\bf k }= \pm \pi$ and ${\bf k }= 0$/$({\bf k}= \pm \pi)$ 
points of the BZ, respectively. Each edge state
splits into two degenerate Kramers doublets, with linear dispersion in a very 
small $k$-region around the crossing and forward and backward mover states with
opposite spin. Since degenerate states are confined at different edges
of the ZGNR, the resulting states with opposite spin are localized in either 
edge. 
Thus two independent spin-filtered edge states are at each boundary of
the ribbon. Hence, for a given energy the ZGNR has four conducting channels 
spatially separated, an extreme of the ribbon contains a forward mover with 
a given spin $\bf S$ and a backward mover with opposite spin, $\bf -S$, and conversely 
for the other extreme of the ribbon. Note, that the energy splitting of the
$\pi$ and $\sigma$ edge states differs almost by a factor of 30.
%of $0.00108eV$ at $k=\pm 0.9\pi$
%while that corresponding to the dispersive $\sigma$-orbital state is of $0.0396eV$
%at $k=\pm 0.1 \pi$. 

\begin{figure}
%\begin{center}
\leavevmode
\includegraphics[width=\columnwidth]{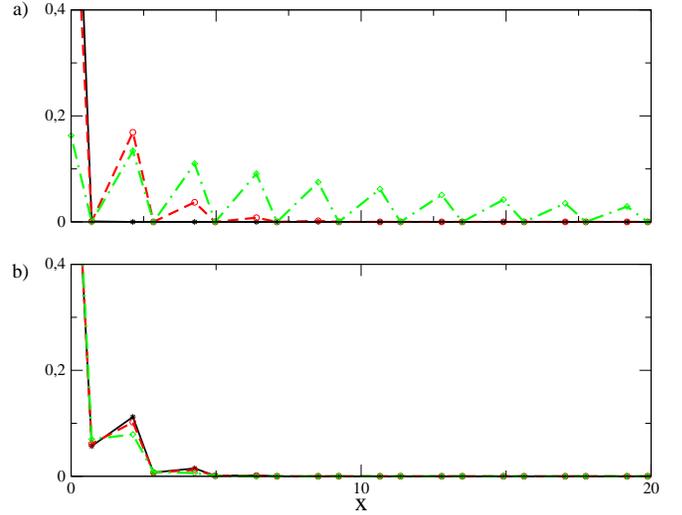}
\caption{(Color on line)Wave function amplitude of the localization versus atomic  coordinates
accross the ribbon width for the two edge states states
represented in Fig. \ref{zedst}. a) for the state at the Fermi level
at $k=\pi$ (black stars), $k= 0.85\pi$ (red circles) and $k=0.7\pi$ 
(green diamonds)
b) for the $\sigma$ derived state at $k=0$(black stars), $k=0.15\pi$ (red circles)
and $k=0.3i\pi$ (green diamonds). Black solid, red dashed and green dott-dashed lines
are guides for the eyes.}
\label{local}

\end{figure}

The localized character of these states can be observed in Figure \ref{local},
where the wave function amplitudes are represented against the
atomic coordinates across the width of the $n=60$ ribbon, for three different
values of $\bf k$ in the BZ. The wave function amplitude of both ${\pi}$ and ${\sigma}$ states is
fully localized in the edge atoms in a large ${\bf k}$ interval around the  
${\bf k}= \pm \pi$ and ${\Gamma}$ points, respectively.
The localization length of the zero-energy state increases with $k$
and eventually, for $\bf k$ close to $\pm  \frac{2\pi}{3}$, where it merges with the bulk
bands, becomes extended. Further, the spatial localization correlates with the 
${\pi}$-${\sigma}$ hybridization induced 
by the SOI. At ${\bf k}=\pm \pi$, where the state is composed of
purely ${\pi}$ orbitals, its wave function is fully localized in the edge
atoms, moving towards the ${\Gamma}$ point, i.e. ${\bf k}= \pm 0.9\pi$,
the localization starts
to decrease toghether with an admixture with ${\sigma}$ orbitals. Therefore,
the presence of the SOI and consequently of the ${\pi}$-${\sigma}$ 
hybridization, slightly reduces the localization length of ${\pi}$ edge states.
Analogously, the ${\sigma}$ edge states show an almost pure 
${\sigma}$ orbital character and a strong localization at the borders of 
the ribbon for both ${\Gamma}$  and ${\bf k }=\pm \pi$  points. The 
presence of a small ${\pi}$ contribution, for ${\bf k}$ values different from 
the high symmetry points, gives rise to the increase of the localization length.

Due to spin-orbit coupling ${\bf S}$ is not longer a good quantum number and 
the eigenfunctions are a linear combination of spin-up and spin-down states.
As stated above, each extreme of the ribbon contains two states, forward and 
backward movers, with opposite spin ${\bf S}$. The calculated expectation value of $<{\bf S}>$
shows that the spin orientation of both ${\pi}$ and ${\sigma}$  edge states
is almost perpendicular to the graphene plane at the time-reversal invariant
${\bf k}= \pi$ and ${\Gamma}$ points. The deviation from the 
perpendicular axis is smaller than 1\%. The orientation axis, although depends
on the magnitude of ${\bf k}$, slightly change in the ${\bf k}$ interval in 
which edge states remain spatially localized. When ${\sigma}$-${\pi}$ 
hybridization becomes relevant and states deslocalize, the in-plane component
has a finite value and the spin orientation axis of extended states forms an 
angle with the graphene plane.
Furthermore, the expectation value of the orbital angular momentum ${< \bf L}>$ for edge states 
is almost zero, indicating the quenching of the orbital angular momentum.

Finally, because of the estimated small value of $\lambda \approx 3$ $meV$,
the corresponding SOI splitting of $\pi$-states is only of the order of 
$0.17 \times 10^{-3}$ $meV$, which gives temperatures in the range of $10^{-3}$ K, in
agreement with previous estimations.
Nevertheless, for the new $\sigma$-derived edge states the 
calculated splitting is much larger, $0.51$ $meV$, which results in
temperatures of the order $\approx 5$ K that, although small, is experimentally attainable 
and thus, the topological protected spin-filtered states of graphene  could
be observed.

\subsection{Klein bearded graphene nanoribbons}

In this termination, as in the case of zigzag edges,
all atoms of one end belong to one sublattice, and those of the
opposite end, to the other sublattice. The Klein's edge could be formed
by bonding an additional C atom to every edge site of the zigzag
ended ribbon \cite{K94}. 
Zigzag and Klein ribbons differ in the number of cut bonds. 

\begin{figure}
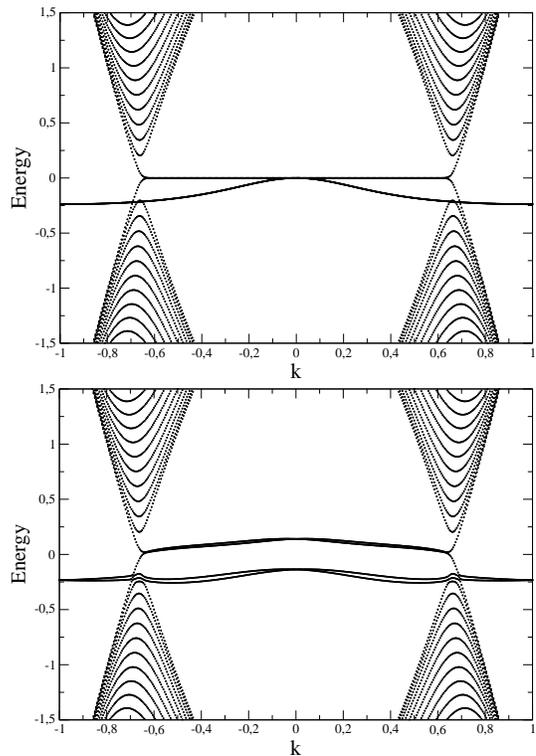


\leavevmode
\includegraphics[width=7.0cm,clip]{fig6a.eps}
\includegraphics[width=7.0cm,clip]{fig6b.eps}
\caption{ Band structure of the bearded edge N=60 ribbon without(with)
SOI left(right). $k$ are in units of $\pi$}
\label{Ker30}
 
\end{figure}

Figure \ref{Ker30}  shows the dispersion relation
of a $n=60$ ribbon with this geometry without and with
the SOI effect.
The Klein boundary produces as well edge states localized
at the ends of the ribbon, with the same behavior of those of the
zigzag nature, although the zero-energy flat bands lie at the central region of the 
BZ \cite{FWNK96,PCNG06}, $-\frac{2\pi}{3} \le {\bf k} \le \frac {2\pi}{3}$. 
Besides the flat zero edge $\pi$ states, there are
two new edge states of $\sigma$ character.
In the absence of SOI, one of the  $\sigma$ states, not shown in the figure, lies well
below the Fermi energy, at around -4 $eV$, 
mixing with the bulk bands, and showing  a weak dispersion along the whole BZ. The other $\sigma$ state, 
lies at zero energy at the ${\Gamma}$ point, but presents energy dispersion, lowering its energy
as $|\bf k|$ increases, lying at  $-0.23$ $eV$ for  $\bf k=\pm \pi$.
The flat bands are of pure $\pi$ character, and are localized in the
outermost atoms at either edge of the ribbon. Also the wave function amplitude
of the the dispersive band of mainly $\sigma$ nature
is strongly localized at the edge atoms, $\approx$ 0.91\% at the extreme atoms.
 The spin-orbit coupling,
besides to induce the splitting into two Kramer's doublets of the edge bands, 
has drastic effects on the energy and dispersion of both $\pi$ and $\sigma$ bands.
While $\pi$ states acquire a small dispersion, a large energy shift and
flattening of the $\sigma$ bands occurs.
These effects  originate from the $\pi$-$\sigma$ hybridization induced by the
SOI and depend on the
value of the coupling strength. For example, 
 the zero energy state was at ${\bf k}=0.2\pi$ pure $\pi$-like
in the absence of SOI term, but has around $1.1\%$ of $\sigma$ components when
the SOI term is considered with the $\lambda=3$ meV coupling.
As shown in Figure \ref{Ker30} bottom, the SOI shifts the edge bands,
inducing a separation   at $\Gamma$, for $\lambda=0.4$ $eV$,  of $276$ $meV$ 
which varies almost linearly with the value of $\lambda$ and therefore it 
reduces to 7.38 $meV$  estimated when  $\lambda=3$ $meV$.

Moreover, the SOI induced energy splitting
of edge states is of $\approx 0.38$ $meV$ and $0.18 \times 10^{-3}$ $meV$ for the
lower and higher energy bands respectively at  $k=\pm 0.2\pi$.
The difference of the splittings is due to the different
orbital compositions of the bands. The energy splitting
of the $\pi$ and $\sigma$ edge states is of the same order
than in the zigzag ribbons. 
%however, for the $\pi$ edge states, the energy splitting  is an order
%of magnitude greater for the bearded termination than for
%the zigzag boundary. 
The dependence of the orbital composition
on ${\bf k}$ explains the different splitting magnitudes
at different regions of the BZ.
%since the $\pi$ flat bands
%appear at different regions of the BZ in both geometries.
The extended bulk subbands states, associated to the intrinsic band structure
of the graphene sheet are, in the presence of spin-orbit coupling,
still doubly degenerate due to the combination of
time reversal and inversion symmetries.
 
The average ${<\bf S>}$ and ${<\bf L}>$ values, analogous to those
of zigzag ribbons, point out the counterpropagation nature of the edge
states and give a spin quantization axes almost perpendicular to the sheet
axes.

\subsection{Armchair graphene nanoribbons}

Armchair edges are formed by homogeneous lines of dimers where  atoms belonging
to the two different sublattices alternate.
The electronic properties
of these ribbons present a strong dependence  on their width, as it happens for zigzag CNTs.
The armchair ended ribbons are metallic when its width $W= (3M-1)d $, 
where $M$ is a positive integer and $d$ is the C-C atom distance. For other values
of $W$ the armchair ribbons (AGNRs) are semiconductors with  direct band gaps at $\Gamma$ which
are inversely proportional to its width \cite{SCL06,E06}. They
follow a family behavior as it is found for zigzag CNTs \cite{CLM09,ZY09},
where metallic behavior is obtained in a sequence of period 3. However, 
zigzag CNTs are metallic for $n=3M$ due to the different  periodic boundary conditions
imposed by the cylindrical geometry.
Ribbons,  with  both metallic  and semiconductor
behavior, of different widths  are calculated
in order to clarify how the SOI affects both electronic structures.

\emph{Metallic AGNRs}

\begin{figure}
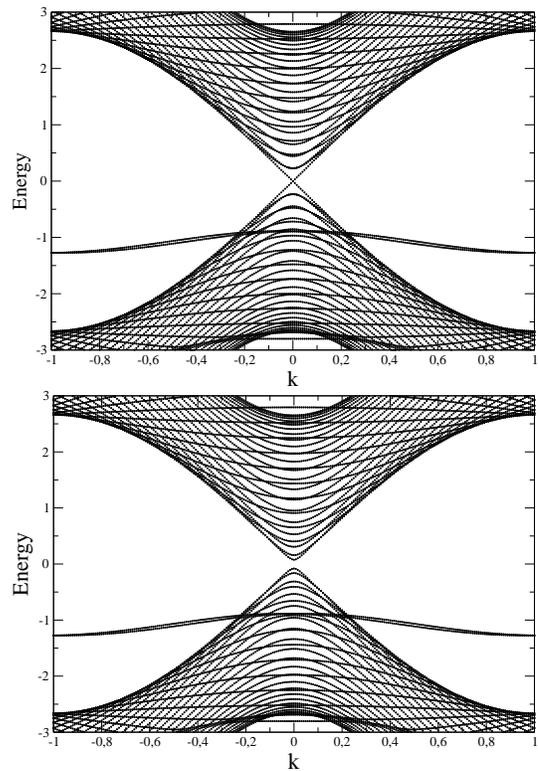


\leavevmode
\includegraphics[width=7.0cm,clip]{fig7a.eps}
\includegraphics[width=7.0cm,clip]{fig7b.eps}
\caption{ Band structure of the armchair  N=62 metallic ribbon (top)
 and N=60 semiconductor ribbon (bottom) with
SOI  $\lambda=0.4$. $k$ are in units of $\pi$}
\label{acr60}

\end{figure}

Armchair ribbons of width $W=(2M-1)d$
present a band structure with two states crossing
at the Fermi  energy at the $\Gamma$ point of the BZ. 
Figure \ref{acr60} shows the band structure of the $n=62$ ribbon calculated with
SOI and $\lambda=0.4$.
In the absence of spin-orbit coupling, the crossing states are only
spin degenerate, have pure $\pi$-orbital character, show linear 
dispersion and their wave functions are extended
throughout the whole ribbon width.
The  inclusion of the SOI term in the Hamiltonian opens a small gap 
at $\Gamma$, for the ribbon {\it n} = 62
is  of $\approx 10^{-4}$ $meV$   for $\lambda=3$ $meV$,
which increases as the width of the ribbon decreases \cite{OIKI08}.
% (for $\lambda=0.8$ the gap is of $\approx 1.4 \times 10^{-3}$eV)
Further, it changes the orbital composition of the linear states
allowing for a  small proportion  about $5\%$, 
of $\sigma$ orbitals  and the pure spin nature is not conserved. 
Nevertheless, the two linear dispersing states remain twofold degenerate
and keep its extended character. These states have been found to persist
in the presence of intrinsic spin-orbit interactions
as spin filtered states  localized on the ribbon edges
within the low-energy Dirac model\cite{ZS07},
In the present model, in order to obtain  localization of these states,
we need unphysical high values of
$\lambda$ of the order of 2.8 eV.

Edge states composed mainly of $\sigma$ orbitals and fourfold 
degeneracy appear below the
Fermi level. They lie at -1.27 eV
at the $k=\pm \pi$ points of the BZ, are dispersive and fully localized
at the edge atoms  of the ribbon. At $k\approx \pm 0.3\pi$ merge with the
bulk bands, but keep its fourfold degeneracy and localized character.
The wave function amplitude is confined to the outermost atoms,
although, at the zone center, localization
occurs at both ends of the ribbon. 
These  $\sigma$-orbital derived states are truly surface states
appearing at the same energy independently of the ribbon width,
as we have confirmed calculating AGNRs of up to $n=122$ and $n=242$.

The $\sigma$ derived edge states, upon SOI inclusion, split into
two doublets, except at the time-reversal invariant ${k=\pm \pi}$ points,
where they remain fourfold degenerate.
The value of the splitting varies with $k$ reaching
a maximun value of $\approx 0.36$ $meV$  at $ k= \pm 0.5 \pi$  for $\lambda=3$ $meV$. 
These states keep the localization
at the boundary atoms, although their orbital composition presents now a small proportion of $\pi$ orbitals.
%con SOI en los k donde antes estan localizados en los dos extremos
%ahora estan en uno altern con mas penetracion
The so-called bulk subbands of the ribbon
mantain always their double degeneracy.

\emph{Semiconductor armchair GNRs.}

The bottom of Fig. \ref{acr60} shows the band structure corresponding
to the semiconductor ribbon N=60, with SOI  $\lambda=0.4$.
The energy gap at  $\Gamma$ is of  $0.156$ eV  almost equal 
to the value obtained without SOI. %for $\lambda=0.0$, $0.1572eV$.
The $\sigma$-orbital derived edge states and bulk related subbands present 
the same features
as in the metallic ribbons. Even the SOI energy splitting of
the Krammers doublets are of the same order than in the previous case.

\subsection{Curvature effects}

Next we analize the effect of curvature. As stated above, in the curved geometry  bond stretching
is not allowed along the ribbon. Even  the distances
between the carbon atoms at the edges are not changed in any of the
different terminations studied. The lattice structure of the ribbon is
isotropically bended along the width, $x$-direction, while
the bonding distances, and therefore the hoppings, between atoms in the
$y$-direction are not modified.

\begin{figure*}
\begin{center}
\leavevmode
\includegraphics[clip,width=\textwidth]{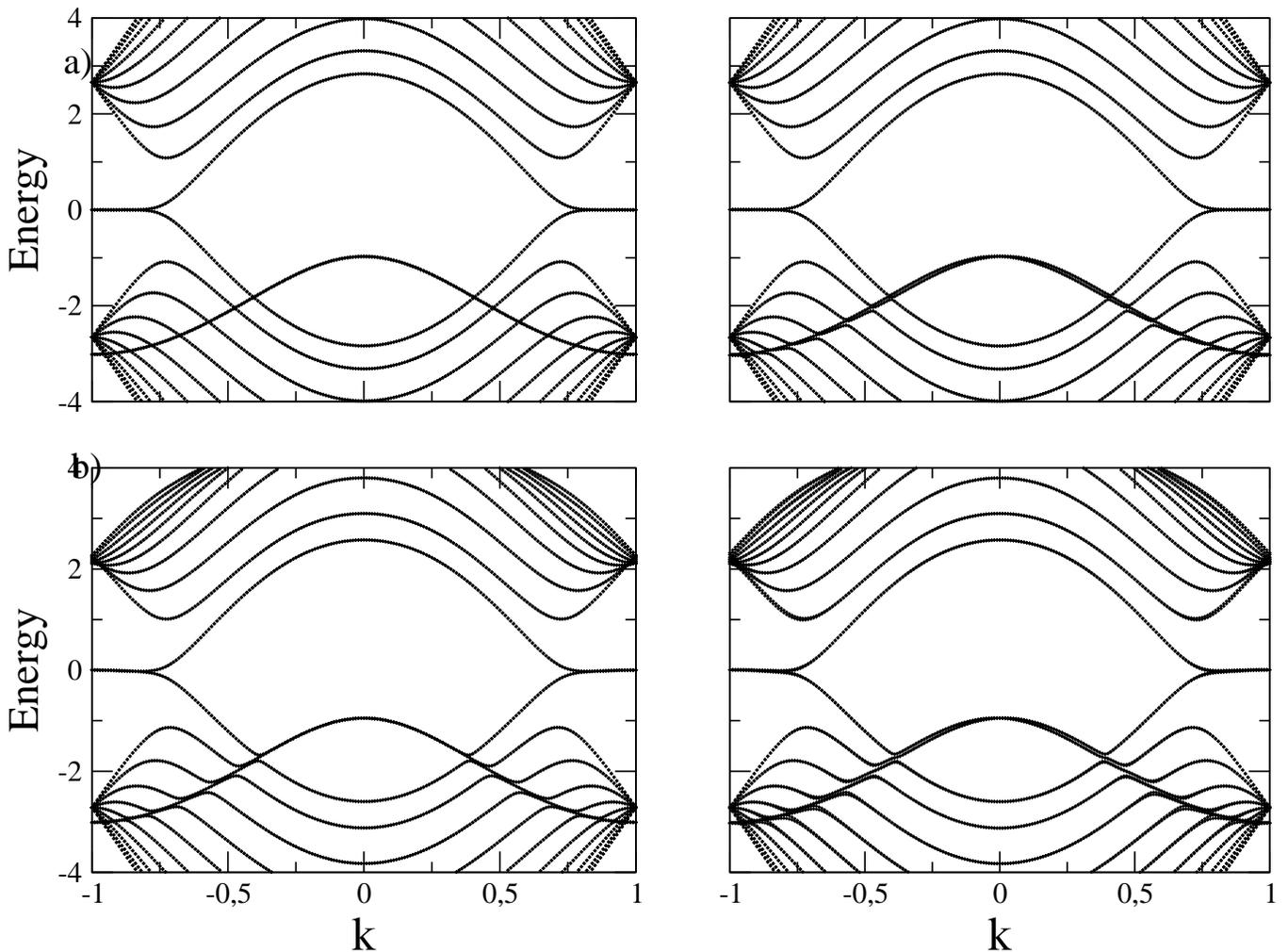}
\caption{Left:Band structure of the N=10 zigzag  nanoribbon
a) flat, b) curved,  
%and c) of the (5,5) CNT,
calculated with $\lambda=0$ (left) and $\lambda=0.4eV$ (right).}
\label{bsrt3}
\end{center}

\end{figure*}

Curvature induces hybridization
between $\pi$ and $\sigma$ orbitals. In fact several works
have shown that  curvature enhances  the SOI effects in CNTs\cite{CLM04,OIKI08,CLM09,JL09}.
The curvature
induced ${\sigma}-{\pi}$ hybridization is related with the
amplification of the SOI effects for small diameter tubes.
However, curvature effects turn out to be weaker for ribbons
than for  CNTs \cite{JCV07,OIKI08}. The cylindrical shape of the tubes impose
periodic boundary conditions that confer symmetries
to the wave functions which do not hold in the open geometries.
As for the flat geometry, a critical value of the width -slightly greater- is needed in
order to avoid the coupling of edge states in the bent ribbons. 
Above the critical value of the
width of the ribbon, the behavior of the edge states found in the flat ribbons
is not affected by the curvature. The localized nature of the edge states
remains upon bending, only in some cases the
localization lenght increases.
In order to illustrate the differences between flat and curved
geometries the electronic structure of the $n=10$ zigzag ribbon,
both flat and bent, are depicted in Figure \ref{bsrt3}.
     
%This result makes it clear
%the topological nature of the edge states in the presence of SOI.
%The curvature effects are also

In the absence of SOI, flat and bent ZGNRs only differ by a small variation
of the subband dispersions. In both geometries edge states, 
$\pi$- and $\sigma$-orbital
derived, appear at similar energies although the orbital composition
of the $\pi$ states is slightly changed with a small admission 
of $\sigma$ orbitals in the curved ribbon. The zero-energy dispersionless
states, in the bent geometry, present a very weak dispersion. %($k=\pm 0.9 \pi$
%is shifted down to $\approx 2meV$ from the zero-energy
%at $k=\pm \pi$). 
Both, the $\sigma$-orbital admission and the dispersion,
increase as $\bf k$ increases, and merge with the bulk-bands at
a $\bf k$ smaller than in the flat geometry. 
%The components of the spin are also changed.
These differences are clear in Figure \ref {bsrt3}, where crossing
of the edge state occurs in the flat
$n=10$ ZGNR band structure while, due to the hybridization, anticrossing 
is observable in the curved case.
SOI effects are similar in  both geometries, although a slight
increase of the splittings appears in the curved ribbon. 
%but smaller than expected (for the $\pi$-state the splittings
%are, at $k=\pm 0.9\pi$ of $0.0011eV$ and $ 0.0039eV$ for the
%flat and curved geometries and for the $\sigma$-state, at
%$k=\pm 0.2 \pi$ of $0.0735eV$ and $0.0836eV$ respectively).

\begin{figure*}

\leavevmode
\includegraphics[clip,width=\textwidth]{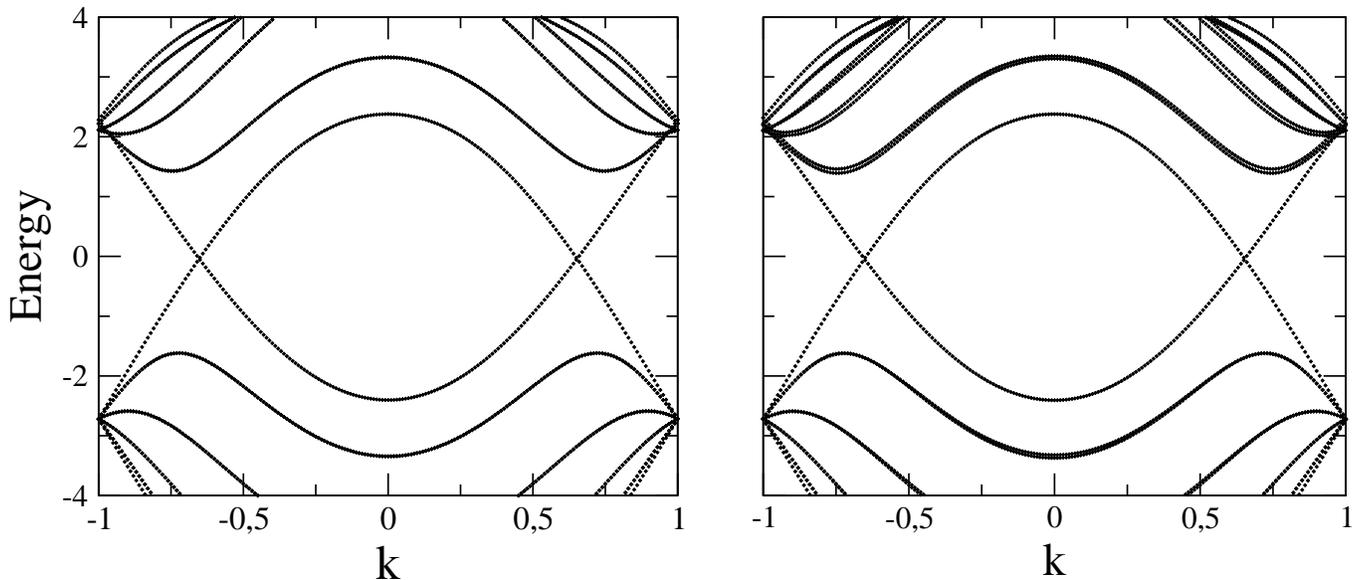}
\caption{Band structure of the armchair (5,5) carbon nanotube
calculated with $\lambda=0$ (left) and $\lambda=0.4eV$ (right).}
\label{tube}

\end{figure*}

In order to compare with the $n=10$ ribbons, 
Figure \ref{tube} represents the band structure of the armchair (5,5) carbon 
nanotube, both without and with SOI. The (5,5) CNT,
could be thought as formed by joining \cite{SCB09} the two borders of the curved
ZGNR $n=10$. Due 
to the breaking of the rotational symmetry, the number of bands
is greater in the ribbon than in the tube.
While the point group of the ZRGN, both flat and bended, is the $C_3$ group of graphene, the
(5,5) tube belongs to the $D_{5d}$ group. The corresponding irreducible
representations
are ten 1D for the $C_3$ and two 1D and four 2D for the $D_{5d}$ \cite {DD65}. Consequently,
considering spin
all the subbands are twofold degenerated in the ZGNR and two
are twofold and four fourfold degenerated in the CNT, 
out of the ten bands
appearing in the energy interval shown in  Figures \ref{bsrt3} and \ref{tube}
(see for example Ref. \cite{CLM04}).
However, in both ribbons and NT, the lower bands crossing at the Fermi level are only spin
degenerate.

The different degeneracy accounts for the different consequences that curvature
has in the ribbons and in the CNTs in the presence of SOI.
As it is well known, SOI removes all degeneracies compatible with time
reversal symmetry and parity. Since ZGNR states are only spin-degenerated,
large splittings comparable to those of the fourfold
degenerate bands of the NT are not observable. Note that the band structure
shown in Figure \ref{bsrt3}  corresponds
to the major possible bended ZGNR, which forms an open cylinder with a
curvature identical to that of the CNT. In fact, curvature effects are
also small in ZGNR in the
presence of the SOI. They reduce  to a slight increase of the energy splitting, see for
 example the second subband. 
Therefore, enhancement of the SOI
strength, analogous to that shown for CNT, does not occur in ribbons.

The effects of curvature on the bearded ribbons are very similar,
inducing changes in the orbital compositions of the edge states as
well as a very weak dispersion in the flat bands. 
At  $\Gamma$ the $\sigma$-orbital
derived edge zero-energy states are shifted down in energy by curvature effects, but  
%$-0.00015eV$
%-0.000065eV$ for the $n=60$ ribbon curved at maximum.
the shift in energy decreases with the ribbon width.
%while the dispersionless $\pi$-derived edge states keep at the Fermi energy, although
%as $k$ moves away from $\Gamma$, the curvature induces a very small dispersion
%and a mixing with $\sigma$ orbitals occurs.
Also, there is an small increase of the energy splittings in the curved geometry when
the SOI is included.
   
On the other hand, the main effect of curvature on the armchair ribbons 
is to induce a small gap at $\Gamma$ on the
metallic ribbons %(of $\approx 0.00057eV$) 
between the two linear
states crossing at the Fermi energy. As well, in both metallic and semiconducting
AGNRs,   
%the fourfold
%degeneracy of the $\sigma$- edge state is lifted by curvature only at $\Gamma$ ($\approx 0.0015eV$).
the SOI induced energy splittings of the $\sigma$ derived edge states are
slightly greater on the curved  than in the flat geometry. 

Curvature also results in a variation of the spin orienation axis and in an 
increase of the orbital angular momentum. 
The expectation value of $<\bf S>$ for the edge states is no longer 
perpendicular to the graphene sheet, but always an in-plane contributation 
turns the spin axis to form an angle with the ribbon plane. Further, $<\bf L>$ 
increases by more than two orders of
magnitude with respect to that of the corresponding flat ribbon. Nevertheless,
for all the ribbons studied it is still much smaller than the orbital moment attributed to CNTs due to
their cylindrical geometry.

In general, the major effect of curvature is on the edge states, 
in the absence of SOI and for flat ribbons with widths greater than 50
chains, the wave functions of the degenerate edge states are
fully localized at either end of the ribbon, thus in the presence of SOI 
they become spin-filtered states. However, for curved ribbons 
the states
localized at the left and right ends interact and
accumulates at both edges, preventing the formation spin conducting chanels.

\section{Discussion and Conclusions}

We have studied the spin-orbit interaction effects on the electronic structure
of graphene nanoribbons taking into account the actual discreteness of the 
lattice. 
The energy subbands associated to the graphene intrinsic electronic bands remain doubly degenerate even when
the SOI term is included in the Hamiltonian. The atomic lattice
structure satisfies spatial inversion symmetry and, in the presence
of time reversal symmetry, the intrinsic spin-orbit coupling
does not break the spin degeneracy \cite{CLM04,KM05}.

The result of the spin-orbit coupling is much more pronounced
in the edge states and particularly in those with a large contribution of 
$\sigma$ orbitals, as it happens in CNTs and bilayer graphene \cite{CLM09,G10}. 
% predominat $\sigma$ character as it
In fact, besides recovering previous results on the edge states originated from $\pi$ 
orbitals, a better understanding of the interplay between the intrinsic 
spin-orbit coupling and the lattice geometry is reached.
It is found that for the flat
zero bands that appear at 
$\frac{2\pi}{3} \le \bf k \le \pi$ and 
$-\frac{2\pi}{3} \le \bf k \le \frac {2\pi}{3}$
in the zigzag and bearded ribbons respectively,
the SOI induced splitting is greater in the regions of the BZ
where  the larger hybridization with $\sigma$ orbitals
takes place. 

The inclusion of $\sigma$ orbitals in the basis set gives rise to 
edge states
missed in one-band calculations. 
Edge states originated from the  $\sigma$-orbitals, 
have been found in the three types of ribbons investigated,
lying below the Fermi energy.  
These states are localized in the
extreme atoms  at different  regions of the BZ (around $\bf k=0$ and $=\pm 
 \pi $). 
As it occurs for the $\pi$-derived states, SOI lifts its  fourfold degeneracy 
except 
at the $\bf k=\pm \pi$ and $\Gamma$ points protected by time reversal symmetry.
A relevant result is that the 
energy splitting is
greater, more than and two orders of magnitude, for the $\sigma$ states than
for the $\pi$ derived states.  This result confirms those found
for CNTs \cite{CLM09} and for a graphene bilayer where, due to the mixing
between $\pi$ and $\sigma$
bands by interlayer hoppings, the spin-orbit coupling is
about on order of magnitude larger than in a single layer \cite{G10}.

While the energy spin-orbit coupling induced
splittings are  of the order of $10^{-3}$ $meV$, equivalent to
T$\approx 0.001$ K for the $\pi$ derived edge bands,   the edge states
originated from $\sigma$ orbitals present splittings of the order of $10^{-1}$ $meV$
{ \it i. e.} T $\approx 4-5$ K. 
These results suggest that $\sigma$ states would be particularly
appropiated to observe the spin currents associated with the spin-filtered
edge states.

In the present work, Fermi energy is taken at zero energy and
neutral graphene is considered, but appropriate chemical doping or external gates
may change this value to tune the energy of the $\sigma$ edge states.
In general, they lie around 1 eV below the Fermi level, although for the 
Klein GNRs the $\sigma$ derived states are only a few meV away from $E_F$.
Therefore, they are easily accesible by applying an external potential, which confers
Klein nanoribons the abbility to exhibit the quantum spin Hall effect.
 
Since curvature is known to 
strongly enhance the spin-orbit induced effects on carbon nanotubes,
we have considered both flat and curved nanoribbons and showed the different
behaviour that open and close boundary conditions impose.

The effects produced by the SOI term and by curvature would 
in principle be similar since both induce $\sigma$-$\pi$ hybridization
and lift degeneracies. Nevertheless, curvature, for the bearded and armchair ribbons, 
lifts some of the degeneracies  at the $\Gamma$ point, while 
SOI does not. 
This result can be understood considering the mapping of the two-dimensional
graphene bands on the axial direction of the ribbons: while
for the zigzag ribbons the Dirac points are projected 
at $\bf k=\pm \frac{2\pi}{3}$ for armchair $K$ and $K'$ points are
mapped to $\Gamma$. The  curvature of the lattice affects
the bands around the $\Gamma$ region of the BZ in the armchair ribbons,
in contrast, for the zigzag termination, the main effects of curvature
occur in the vicinity of $k=\pm \frac{2\pi}{3}$.   
Analogously in CNTs curvature induces a gap in
the primary metal $n=3q$ zigzag tubes while armchair CNTs
are metallic.

The interplay between curvature and SOI effects in nanoribbons 
are not as significant as in CNTs \cite{OIKI08}
The boundary conditions imposed by the cylindrical
shape of CNTs confer rotational symmetry which,
alongside to the bipartite nature of the honeycomb lattice,
is in the origin of the unusual electronic properties of
CNTs. Orbital magnetic moments 10-20 times larger than
the Bohr magneton have been observed in suspended
CNTs \cite{MYSM04}, attributed to semiclassical 
electron orbits encircling the tube circumference, 
whose diameter is much larger than the radii of the
atomic orbitals \cite{MYSM04}. 
This behavior does not occur in the bent ribbons, where no rotational
symmetry holds.
The spin-orbit coupling effect in bent ribbons is mostly due to the
increased $\pi$-$\sigma$ hybridization. Thus, both curvature and SOI
contribute to the increase of the orbital angular momentum and to deviate
the spin orientation 
from the perpendicular direction to the graphene sheet.
Finally, it is found that curvature does not break the chiral symmetry
of the edge states and its effects on the localization
reduce to a  small increase of the localization length. 
%(At $\Gamma$ the zero state is degenerate eight times in the
%flat ribbon: the maximum curvature splitts in two fourfold
%degenerate one at zero and the other at $-0.000065eV$,
%while SOI, $\lambda=0.4$ splitts 0.276eV).
%In the curved geometry the splittings induced by
%SOI are increased $\approx 0.0005eV$ and $\approx 0.002 eV$ 
%for the $\sigma$ and $\pi$ bands respectively. 

In summary, we estimate the magnitude of the intrinsic spin-orbit coupling 
to be of $\approx$ 3 meV and although small, we have demostrated that, under certain geometries, and 
experimental conditions the Quantum Spin Hall state predicted in graphene 
should be
experimentally observed in nanoribbons at temperatures of the order of 3-4 K.
Further, the spin-filtered edge-states are robust and unaffected by curvature.
Ours results highlight the importance of symmetry to 
understand the
spin-orbit coupling effects in graphene nanoribbons.

%%%%%%%%%%%%%%%%%%%%%%%%%%%%%%%a

%%%%%%%%%%%%%%
\section{acknowledgments}
The authors thank L. Brey, L. Chico and F. Guinea for discussions. 
This work has been partially supported by the Spanish DGES under
grants MAT2006-05122, FIS2008-00124 and MAT2009-14578-C03-03.

\bibliography{ribsobibl}
\end{document}